\documentclass[11pt,twoside,a4paper,cmspaper,final,collab]{cms-tdr}
\def\svnVersion{9a9efe9}\def\svnDate{2025/12/27}\def\cmsCernNoTag{CERN-EP-2025-046}\def\cmsCernDate{\today}\def\cmsMessage{Published in Physics Letters B as \href{http://dx.doi.org/10.1016/j.physletb.2025.140120}{\doi{10.1016/j.physletb.2025.140120}.}}
\begin{document}\cmsNoteHeader{HIN-23-006}

\providecommand{\PGgst}{{\HepParticle{\PGg}{}{\ast}}\xspace} 
\newcommand {\roots}    {\ensuremath{\sqrt{s}}}
\newcommand {\rootsNN}  {\ensuremath{\sqrt{\smash[b]{s_{_{\mathrm{NN}}}}}}\xspace}
\newcommand {\deta}     {\ensuremath{\Delta\eta}\xspace}
\newcommand {\dphi}     {\ensuremath{\Delta\phi}\xspace}
\newcommand {\pp}    {\ensuremath{\Pp\Pp}\xspace}
\newcommand {\PbPb}  {\ensuremath{\text{PbPb}}\xspace}
\newcommand {\ptt}    {\ensuremath{\pt^\text{ch}}\xspace}
\newcommand {\etat}    {\ensuremath{\eta^\text{ch}}\xspace}
\newcommand {\phiz}    {\ensuremath{\phi^{\PZ}}\xspace}
\newcommand {\phit}    {\ensuremath{\phi^\text{ch}}\xspace}
\newcommand {\yz}    {\ensuremath{y^{\PZ}}\xspace}
\newcommand{\ptz} {\ensuremath{\pt^{\PZ}}\xspace}
\newcommand{\dphitz}{\ensuremath{\Delta\phi_{\mathrm{ch,Z}}}\xspace}
\newcommand{\dytz}{\ensuremath{\Delta y_{\mathrm{ch,Z}}}\xspace}%
\newcommand{\zay}{normalized associated yield\xspace}
\newcommand{\hybrid}{\textsc{Hybrid}\xspace}
\newcommand{\jewel}{\textsc{Jewel}\xspace}
\newcommand{\pyquen}{\textsc{Pyquen}\xspace}
\newcommand{\colbt}{\textsc{Co-LBT}\xspace}

\cmsNoteHeader{HIN-23-006}

\author[cern]{The CMS Collaboration}

\title{Evidence of medium response to hard probes using correlations of \texorpdfstring{\PZ}{Z} bosons with hadrons in heavy ion collisions}

\author[inst1]{Yi Chen, Yen-Jie Lee, Pin-Chun Chou, Hannah Bossi, Chris McGinn, Gian Michele Innocenti}

\date{\today}

\abstract{The first measurement of pseudorapidity and azimuthal angle distributions relative to the momentum vector of a \PZ boson for low transverse momentum (\pt) charged hadrons in lead-lead (\PbPb) collisions is presented. By studying the hadrons produced in an event with a high-\pt \PZ boson ($40 < \pt < 350\GeV$), the analysis probes  how the quark-gluon plasma (QGP) medium created in these collisions affects the parton recoiling opposite to the \PZ boson. Utilizing \PbPb data at a nucleon-nucleon center-of-mass energy $\rootsNN = 5.02\TeV$ from 2018 with an integrated luminosity of 1.67\nbinv and proton-proton (\pp) data at the same energy from 2017 with 301\pbinv, the distributions are examined in bins of charged-hadron \pt. A significant modification of the distributions for charged hadrons in the range $1 < \pt < 2\GeV$ in \PbPb collisions is observed when compared to reference measurements from pp collisions. The data provide new information about the correlation between hard and soft particles in heavy ion collisions, which can be used to test predictions of various jet quenching models. When compared with theoretical predictions, the results are consistent with expectations of a hydrodynamic wake created when the QGP is depleted of energy by the parton propagating through it. Therefore, this Letter presents the first evidence of probe-induced energy depletion and the resulting response by the QGP medium. 
}

\hypersetup{
pdfauthor={CMS Collaboration},
pdftitle={Evidence of medium response to hard probes using correlations of Z bosons with hadrons in heavy ion collisions},
pdfsubject={CMS},
pdfkeywords={CMS, heavy ions}}

\maketitle

\section{Introduction}
Quantum chromodynamics (QCD) predicts that a state known as the quark-gluon plasma (QGP), which consists of deconfined quarks and gluons, is formed at high temperatures and high density~\cite{PhysRevLett.34.1353, Karsch:1995sy}. The QGP is created in relativistic heavy ion collisions~\cite{Busza:2018rrf} and has been the focus of high-energy nuclear physics experiments at BNL RHIC~\cite{STAR:2005gfr,PHENIX:2004vcz,PHOBOS:2004zne,BRAHMS:2004adc} and the CERN LHC~\cite{ALICE:2022wpn, CMS:2024krd}. During heavy ion collisions, parton scattering with high momentum transfer occurs earlier than the formation of the QGP. Therefore, such partons can serve as tomographic probes of the plasma ~\cite{PhysRevD.27.140}. These partons lose energy by interacting strongly with the QGP medium, an effect which results in ``jet quenching''~\cite{Gyulassy:1990ye}. Experimental measurements of high-transverse-momentum (\pt) hadrons~\cite{Adams:2003kv, Adare:2008qa, Abelev:2012hxa, Aad:2015wga, CMS:2012aa, Khachatryan:2016odn} and jets~\cite{Chatrchyan:2011sx, Aad:2010bu, ATLAS:2014cpa, Adam:2015doa, Khachatryan:2016jfl, Adam:2015ewa, Adamczyk:2016fqm} formed by the fragmentation of high-momentum partons have provided evidence establishing the jet-quenching phenomenon. However, the details of the jet quenching mechanism remain elusive. Processes such as collisional energy loss, where partons lose energy through interactions with medium constituents, and medium-induced radiation in perturbative QCD (pQCD)-based models, reduce the momentum of the parton as it traverses the QGP. In anti-de Sitter/conformal field theory (AdS/CFT)-based models, a drag force, scaled to account for the differences between QCD and infinitely-color-charged field theory, similarly reduces the energy of the parton when passing through the medium~\cite{Casalderrey-Solana:2014bpa}. Additionally, medium recoils and the response induced by the hard probes can contribute to producing low-\pt particles linked with the high-energy parton (recent reviews can be found in Refs.~\cite{Cao:2020wlm,Apolinario:2022vzg,Cao:2022odi}).
 Since both contributions are associated with the QGP temperature scale on the order of hundreds of \MeV, separating them using their \pt distributions is difficult. This highlights the need for more observables that are sensitive to the different contributions to the jet-quenching mechanism and for determining the relevance of medium recoil and medium response.

Unlike strongly interacting particles, electroweak bosons do not significantly interact with the QGP~\cite{Chatrchyan:2012vq, Chatrchyan:2012nt,Chatrchyan:2014csa,ATLAS:2015rlt,CMS:2020oen,CMS:2021kvd}. Therefore, measurements of jets produced in the same hard scattering as these bosons offer a controlled configuration for the initial hard scattering~\cite{Kartvelishvili:1995fr, Wang:1996yh, Wang:1996pe, Brewer:2021hmh}. The \pt of the electroweak boson reflects the initial energy of the associated parton before any medium-induced energy loss occurs~\cite{Dai:2012am, Kang:2017xnc}. Among the electroweak bosons, \PZ bosons offer the most precise means to identify the hard-scattering scale for $\PZ+$jet events. Although W bosons have a larger production cross section, the presence of a neutrino in their leptonic decay products complicates the boson \pt measurement. Additionally,  compared to the related measurements using isolated photons~\cite{Chatrchyan:2012vq,ATLAS:2015rlt,Sirunyan:2018qec,Sirunyan:2018ncy,ATLAS:2024prm}, \PZ bosons do not require a complex isolation procedure that could introduce measurement biases for particle production in their vicinity. In this analysis, \PZ bosons are reconstructed via their dimuon decay channel, which has negligible background contamination~\cite{CMS:2021kvd} when observed using the CMS detector~\cite{CMS:2017eqd, Brewer:2021hmh}.
 These advantages enable the experimental verification of the presence of a negative diffusion wake, where the hard-scattered parton pushes the QGP medium forward, leaving a depleted region on the \PZ boson side. Furthermore, these traits allow for the study of medium response and recoil in a phase space far away from the outgoing parton~\cite{CMS:2021otx, Yang:2021qtl,Yang:2022nei,ATLAS:2024prm} by examining the charged-hadron spectra in the vicinity of the \PZ boson. Looking in the \PZ boson direction avoids complications faced by previous searches (e.g., Ref.~\cite{CMS:2021otx,ATLAS:2020wmg}) that examined the region opposite the \PZ boson, where other jet quenching effects contribute in addition to the medium response.
 
Moreover, examining the angular distributions of hadrons relative to the \PZ boson as a function of hadron \pt allows the determination of where the medium response and recoil effects are the most significant. The study of high-\pt charged hadrons (high $\ptt$) facilitates searches for the large-angle scattering between high-\pt partons and QGP constituents~\cite{DEramo:2012uzl, He:2015pra}.

This Letter presents the first measurement of the \PZ boson-hadron two-particle angular correlation function in bins of hadron \pt. Charged-hadron spectra in events containing high-\pt \PZ bosons are studied as functions of the azimuthal angle difference between the \PZ boson and the charged hadron ($\dphitz\equiv \phiz-\phit$), as well as their rapidity difference ($\dytz\equiv\yz-\etat$), where $\yz$ is the rapidity of the \PZ boson and $\etat$ is the pseudorapidity of the charged hadron. This study utilizes lead-lead (\PbPb) data at a nucleon-nucleon center-of-mass energy $\rootsNN = 5.02\TeV$ recorded in 2018 by CMS with an integrated luminosity of $1.67 \pm 0.03\nbinv$, as well as proton-proton (\pp) data acquired in 2017 with an integrated luminosity of $301 \pm 6\pbinv$ at the same energy~\cite{CMS:2025rzq}.  Tabulated results are provided in the HEPData record for this analysis~\cite{hepdata}.

\section{The CMS detector}

The central feature of the CMS detector~\cite{Chatrchyan:2008zzk} is a superconducting solenoid of 6\unit{m} internal diameter, providing a magnetic field of 3.8\unit{T}. Within the solenoid volume are a silicon pixel and strip tracker, a lead tungstate crystal electromagnetic calorimeter (ECAL), and a brass and scintillator hadron calorimeter. Hadron forward (HF) calorimeters extend the $\eta$ coverage up to $\abs{\eta}=5.2$. For \PbPb events, the amount of energy measured in HF is used to determine the degree of overlap (``centrality'') of the two colliding nuclei~\cite{Chatrchyan:2011sx}. Muons are measured in gas-ionization detectors located outside the solenoid.

\section{Monte Carlo simulation}
The \MGvATNLO v2.4.2~\cite{Alwall:2014hca} Monte Carlo (MC) event generator at next-to-leading order (NLO) interfaced with \PYTHIA~8.212~\cite{Sjostrand:2014zea} with tune CP5~\cite{CMS:2019csb} are used to simulate signal Z+jets events. In the case of \PbPb collisions,  samples are created by embedding the \MGvATNLO{}+\PYTHIA signal events in minimum-bias heavy ion events modeled using the {\HYDJET}1.9 MC event generator \cite{Lokhtin:2005px}. These events are then propagated through the CMS apparatus using the \GEANTfour toolkit \cite{geant4}, and are used to evaluate reconstruction and selection efficiencies, perform calibrations, and study the background. 

\section{Global-event and physics object selection}
The \PZ boson event samples are selected online using dedicated lepton triggers \cite{Khachatryan:2016bia}. The triggers on $Z \to \PGmp \PGmm$ events require one muon with $\pt > 12\GeV$ and $\abs{\eta}< 2.4$ \cite{Khachatryan:2016bia}. Noncollision events, such as beam-gas interactions, are removed in the data analysis following the procedure in Ref.~\cite{Khachatryan:2016odn}. The average pileup, defined as the mean number of additional collisions within the same bunch crossing, is 2 in \pp collisions  and smaller than 0.01 in \PbPb collisions. In the \pp analysis, events with a single primary vertex, formed by at least two tracks, with a distance from the center of the nominal interaction region of less than 25 cm along the beam axis, are selected to reject events with pileup. Centrality quantifies the overlap of the two colliding nuclei and is related to the initial energy density in the system. It is determined based on percentiles of the total energy deposited in the HF calorimeters, representing the fraction of the total inelastic hadronic cross section, starting at 0\% for the most central collisions \cite{Chatrchyan:2011sx}. The results for \PbPb collisions are presented in three centrality intervals: 0--30, 30--50, and 50--90\%.

The muon selection requires reconstructed track segments in at least two muon detector planes and a good-quality fit when connecting them to tracker segments \cite{CMS:2024qgs}. For both \pp and \PbPb data, muons must have $\pt > 20\GeV$ and fall within the acceptance of the muon detectors, \mbox{$\abs{\eta}<2.4$}. Muon reconstruction efficiencies, found using MC simulations, are corrected according to the their data-to-MC differences, obtained using the ``tag-and-probe" method~\cite{CMS:2011aa,CMS:2024qgs}.

The charged-hadron track reconstruction used in both \pp and \PbPb collisions is described in Ref.~\cite{Chatrchyan:2014fea}. Corrections for tracking efficiency, detector acceptance, and misreconstruction rate are obtained following the procedure outlined in Ref.~\cite{Khachatryan:2016odn}. The selection criteria are identical to those described in Ref.~\cite{Khachatryan:2016odn} for the \pp and \PbPb data.

\section{Analysis technique and corrections}

The \PZ boson candidates are identified using an oppositely-charged muon pair, with a reconstructed invariant mass in the interval 60--120\GeV and $40<\ptz<350\GeV$. Here, contributions from virtual photons decaying into an oppositely-charged muon pair ($\PGgst\to \PGmp\PGmm$) are also included in the signal definition. In events with more than one \PZ boson candidate, which comprises approximately 0.2\% of events, the highest \pt boson is used for the correlation analysis. About 2000 events pass these selection criteria in the \PbPb analysis. Muon pairs are corrected for losses due to detector acceptance, reconstruction efficiency, muon identification, and trigger selection criteria~\cite{Chatrchyan:2012xi}. Each \PZ boson candidate is correlated with all tracks in the same event that have $\ptt > 1\GeV$ and $\abs{\etat} < 2.4$. To avoid including the muons from the \PZ  boson candidate decay, each track used in the correlations is required to fall outside a cone  around the direction of a muon from the \PZ boson decay. The cone radius, defined as $\sqrt{\smash[b]{{(\deta)}^2 + {( \dphi)}^2}}$, is required to be greater than 0.0025 radians. This is the smallest value that ensures no significant muon contamination, as determined from simulations. The simulation gives a good description of the resulting invariant mass distributions for reconstructed \PZ bosons~\cite{CMS:2017eqd}.  
 
In order to focus on the shape of the charged hadron spectrum associated with the \PZ~boson, the average over the $N_{\text{ch}}$ distribution, $\langle  N_\text{ch}\rangle$ where $N_{\text{ch}}$ is the number of associated charged hadrons, needs to be subtracted. Therefore, the averaged normalized associated yield  $\Delta \langle N_\text{ch}\rangle$, is defined as
\ifthenelse{\boolean{cms@external}}
{
\begin{multline}
\Delta \langle N_{\text{ch}} \rangle (\dphitz,\dytz) \\ \equiv \frac{1}{N_\PZ} \sum_{i=1}^{{N_\PZ}} (N_{\text{ch},i} (\dphitz,\dytz) - \langle N_{\text{ch}} \rangle),
\label{eq1}
\end{multline}
}{
\begin{equation}
\Delta \langle N_{\text{ch}} \rangle (\dphitz,\dytz)  \equiv \frac{1}{N_\PZ} \sum_{i=1}^{{N_\PZ}} (N_{\text{ch},i} (\dphitz,\dytz) - \langle N_{\text{ch}} \rangle),
\label{eq1}
\end{equation}
}
where $N_{\PZ}$ is the number of leading \PZ boson candidates, defined as the highest \pt \PZ boson in the event. In order to extract the shape $\Delta \langle N_\text{ch}\rangle$ of the charged hadrons that are associated with the parton recoiling opposite the Z boson, an event-mixing scheme is employed to subtract the substantial background arising from the many unrelated multiparton interactions in heavy ion collisions. The mixed-event correlation function is calculated by pairing the \PZ boson from one event with charged hadrons from a different \PZ boson event within the sample. The impact of not applying any further event criteria (beyond the presence of a \PZ boson) in selecting events for this mixing was evaluated with MC closure studies and was found to be negligible. To improve the statistical accuracy, this procedure is repeated using 10 other events for each event in the sample. In this context,  $S(\dphitz,\dytz)$ and $B(\dphitz,\dytz)$ denote the same-event and mixed-event correlation functions, respectively. The relationship between $\Delta \langle N_{\text{ch}}\rangle$ and these functions can be written as:
\ifthenelse{\boolean{cms@external}}
{
\begin{multline}
\Delta \langle N_{\text{ch}} \rangle (\dphitz,\dytz) \\= S(\dphitz,\dytz) - B(\dphitz,\dytz).
\end{multline}
}{
\begin{equation}
\Delta \langle N_{\text{ch}} \rangle (\dphitz,\dytz) = S(\dphitz,\dytz) - B(\dphitz,\dytz).
\end{equation}   
}

This method, applied to both $\Pp\Pp$ and \PbPb analyses, subtracts the underlying event and $\langle N_{\text{ch}} \rangle$ contributions, isolating the medium-induced modifications. Note that no unfolding corrections for detector effects in  \dphitz (\dytz) are applied in this case, as the bin width is much larger than the resolution. Results for positive and negative \dphitz (\dytz) are merged to improve statistical precision. These symmetrized results, which are effectively a correlation function in bins of  $\abs{\dphitz}$ ($\abs{\dytz}$), are then scaled by a factor of 0.5 to ensure that the total integrated yield accurately reflects the relative excess or depletion observed in the data, thereby allowing direct comparison with theoretical predictions. To make a clearer visual presentation of the data, the results are ``reflected," i.e., the same values are plotted at both positive and negative \dytz, while for \dphitz data points are reflected around both 0 and $\pi$.

\section{Systematic uncertainties}
Several variations in the analysis are considered to account for uncertainties related to tracking efficiency, muon identification efficiency, muon reconstruction efficiency, \pp pileup, and \PbPb background subtraction. Systematic uncertainties are either evaluated as the difference between the nominal results and those obtained with varied settings or are based on the differences in various quantities between data and simulation. 

The uncertainty related to the tracking efficiency is estimated by comparing the track reconstruction efficiency between data and simulation~\cite{Khachatryan:2016odn}. An efficiency uncertainty of 5.0 (2.4)\% per charged-hadron track is quoted as the systematic uncertainty in the \PbPb (pp) analysis. Muon triggering and reconstruction efficiencies are varied within the uncertainties of their data-to-MC differences, as determined with the tag-and-probe method. For systematic uncertainties associated with pileup in \pp collisions, nominal distributions are compared to those from events without the single-primary-vertex requirement. The centrality calibration is varied to calculate the effect of the minimum-bias event selection efficiency of the HF calorimeters~\cite{CMS:2021kvd,CMS:2024krd}. In order to investigate the uncertainty in the \PbPb background subtraction, an MC closure test was performed, which resulted in a negligible non-closure and corresponding uncertainty. The uncertainties are then summed in quadrature to obtain the final systematic uncertainties, and are significantly smaller than the statistical uncertainty. When calculating the systematic uncertainties in the difference between \pp and \PbPb spectra, the systematic uncertainties in the two systems are assumed to be uncorrelated. The dominant systematic uncertainties are those associated with pileup and the centrality calibration in the \pp and \PbPb analyses, respectively. The magnitude of the total relative systematic uncertainty varies with centrality and \ptt bin. For \dphitz, the systematic uncertainty is largest for the lowest-\ptt interval for 30--50\% centrality with a total relative systematic uncertainty of $\sim$75\%. For \dytz, the systematic uncertainty is largest at lower \ptt and 50--90\% centrality, with a total relative systematic uncertainty of $\sim 40\%$. However, in all regions of phase space, the statistical uncertainty is larger than the systematic uncertainty.

\section{Results}

\begin{figure*}[ht!]
    \centering
    \includegraphics[width=0.7\textwidth]{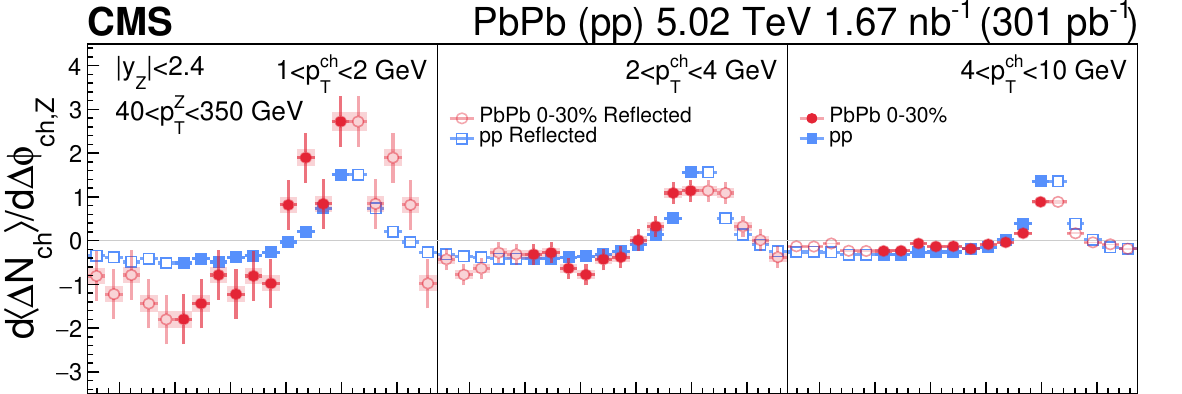}
    \includegraphics[width=0.7\textwidth]{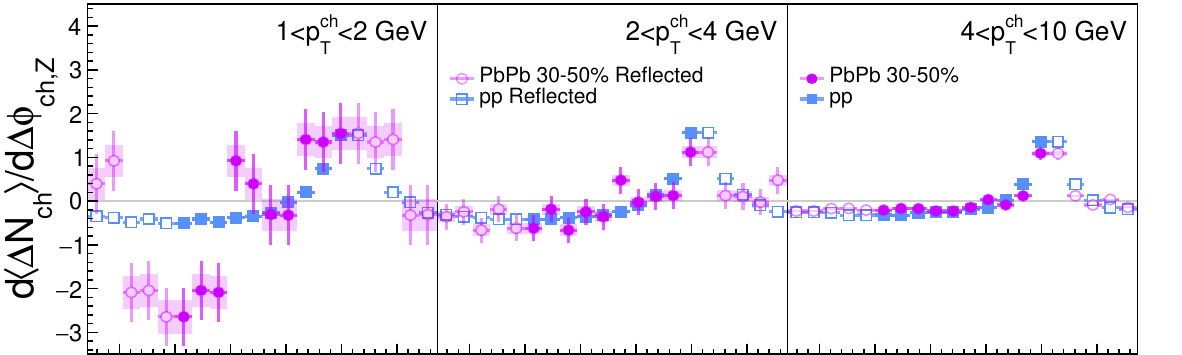}
    \includegraphics[width=0.7\textwidth]{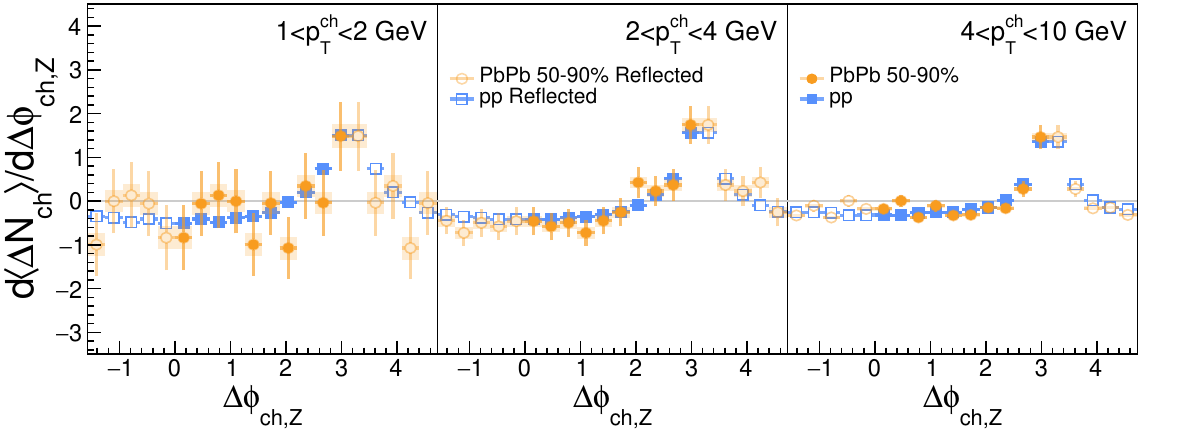}
    \caption{The $\dphitz$ spectra for events with Z boson $\ptz > 40\GeV$ in \pp and \PbPb collisions. The filled circles (squares) are the \PbPb (\pp) data and the open circles (squares) are reflected data. The \pp data are identical across panels with different centrality intervals.  The vertical bars and shaded boxes represent the statistical and systematic uncertainties, respectively. The results are presented in centrality intervals of 0--30\% (top row, red), 30--50\% (middle row, purple), and 50--90\% (bottom row, yellow) and in the charged-hadron \ptt intervals of 1--2 (left column), 2--4 (middle column) and 4--10\GeV (right column).}
    \label{fig:result-dphi}
\end{figure*}

The distributions of the azimuthal angle difference, \dphitz, between the \PZ boson and charged-hadron tracks in three \ptt intervals are shown in Fig.~\ref{fig:result-dphi}. In \pp collisions, the distribution features a jet-side peak at $\dphitz \sim \pi$, while on the \PZ boson side, defined as the region close to the \PZ boson ($\dphitz\sim 0$), the \zay is flat and becomes negative. To investigate the centrality dependence of the medium-induced modification, the results from \PbPb data are studied in three centrality classes: 0--30, 30--50, and 50--90\%. The fully-corrected \PbPb data are then compared to the reference \pp data.

In the 0--30\% centrality class, significant modifications are observed in the normalized distributions compared to the \pp reference in the highest- and lowest-\pt intervals. A more pronounced modulation in the \dphitz distribution is observed for low-\pt tracks within the $1<\pt<2\GeV$ range, consistent with the expected response from parton-medium interactions.
For the intermediate-\pt range of $2<\pt<4\GeV$, the \PbPb data show a closer resemblance to the \pp reference than is the case for the lowest-\pt charged hadrons. 
Significant suppression of high-\pt particles in \PbPb is observed on the jet-side for $4<\pt<10\GeV$, especially when $\abs{\dphitz} > \pi/2$. The difference between the \PbPb and \pp data in the jet side diminishes for more peripheral collisions. The \pp data and the \PbPb data at 50--90\% centrality are consistent with each other.

\begin{figure*}[ht!]
    \centering
    \includegraphics[width=0.7\textwidth]{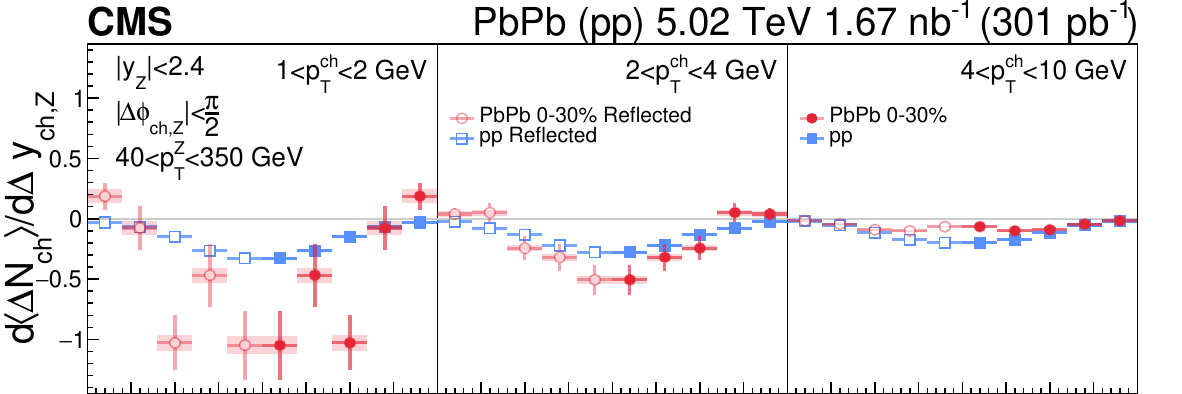}
    \includegraphics[width=0.7\textwidth]{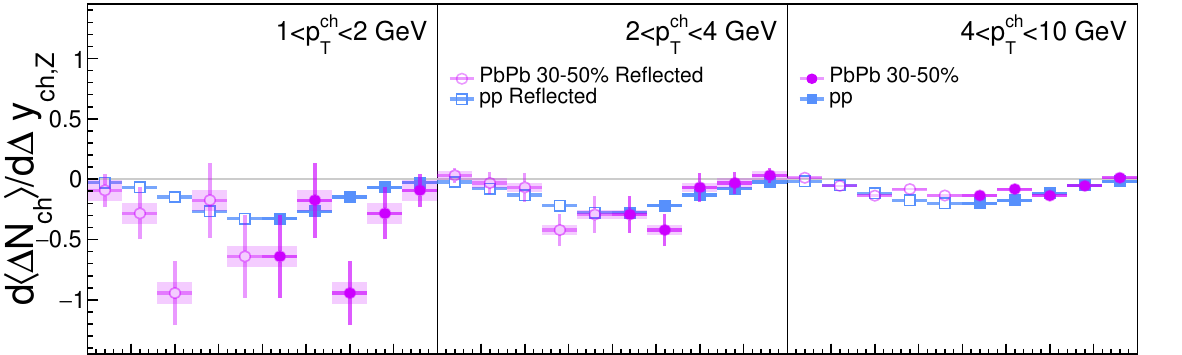}
    \includegraphics[width=0.7\textwidth]{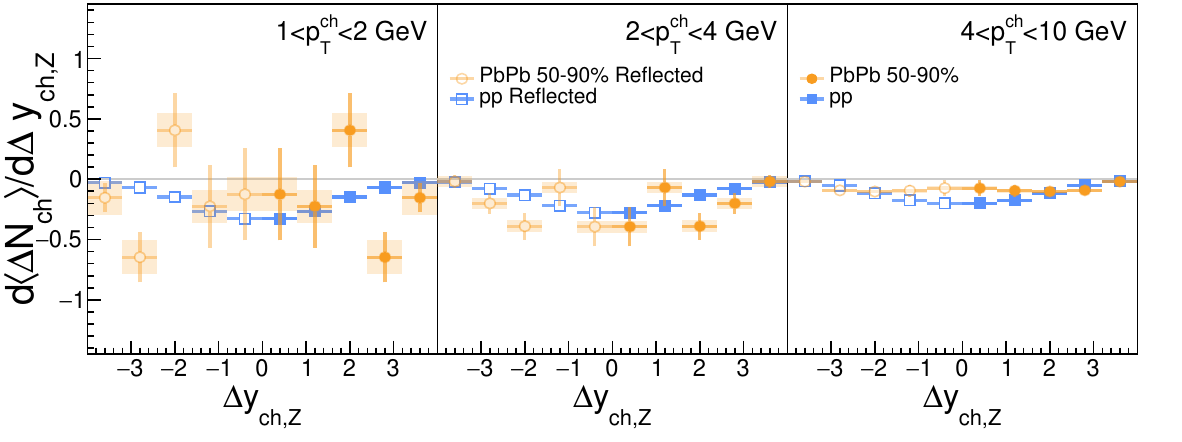}
    \caption{The $\dytz$ spectra in the Z boson side ($\abs{\dphitz}<\pi/2$) for events with Z boson $\ptz > 40\GeV$ in \pp and \PbPb collisions. The filled circles (squares) are the \PbPb (\pp) data and the open circles (squares) are reflected data. The vertical bars and shaded boxes represent the statistical and systematic uncertainties, respectively. The results are presented in centrality intervals of 0--30\% (top row, red), 30--50\% (middle row, purple), and 50--90\% (bottom row, yellow) and in the charged-hadron \ptt intervals of 1--2 (left column), 2--4 (middle column) and 4--10\GeV (right column).  The \pp data are identical across panels with different centrality intervals. }
    \label{fig:result-dy-z}
\end{figure*}

The correlations as a function of the rapidity difference between the charged hadrons and the \PZ boson, $\dytz$, are presented in Fig. \ref{fig:result-dy-z}. To focus on the Z direction, the rapidity difference distributions include only particles with $\dphitz<\pi/2$. Because of this azimuthal angle selection, these spectra do not average to zero. On the \PZ boson side, the correlations are sensitive to the medium-recoil effect, which would appear as a dip structure at $\dytz \sim 0$ relative to the \pp reference. As shown in all panels in Fig. \ref{fig:result-dy-z}, the \zay for \pp data on the \PZ boson side is smaller than zero. The same measurement is also performed with \PbPb data. At low \ptt ($1<\pt<2\GeV$), the 0--30\% centrality \PbPb data are below the \pp reference, with a significance greater than $3\sigma$ using a $\chi^2$ test between the \pp and \PbPb data points. This difference diminishes for $2<\pt<4\GeV$. 
Bin-to-bin correlations in the systematic uncertainties have only a minor effect on the calculation of $\chi^2$ and are therefore neglected. For the highest \pt bin ($4<\pt<10\GeV$), the 0--30\% centrality \PbPb data are higher than the \pp reference. This phenomenon is consistent with expectations of jet quenching, likely driven by the suppression of the jet-side yield and potential large-angle parton-medium scattering. In the most peripheral \PbPb collisions (centrality 50--90\%), the difference between the \pp and \PbPb data reduces at the low \ptt.

The \PbPb data are also compared to jet quenching models. The ``\PYTHIA-quenched" (\pyquen 1.5.3) model approximates radiative and collisional energy loss without strictly enforcing local energy-momentum conservation~\cite{Lokhtin:2011qq}. The model can be used to predict the shower of high-energy partons traversing the QGP, resulting in medium-induced modifications to jet structures and particle correlations. This model predicts that the \PZ-boson-hadron correlation is influenced by collisional and radiative effects induced by the medium, leading to substantial changes in the jet-fragmentation function.

The ``jet evolution with energy loss" (\jewel 2.2.0) model simulates the interaction between jets and the QGP with pQCD-based calculations, incorporating both radiative and collisional energy losses~\cite{Zapp:2008gi,Zapp:2012ak,Zapp:2013vla}. It predicts jet modifications due to the medium, including the recoil of medium particles. This model can operate in two configurations: with recoil, which conserves energy and momentum, and without recoil, which does not conserve these quantities but allows for focusing on the quenched parton shower~\cite{Zapp:2013vla,KunnawalkamElayavalli:2017hxo}. According to this model, medium recoils significantly enhance the low-\pt \zay on the jet side. In addition, the normalized yield around the Z boson is predicted to be suppressed by ``medium holes,'' which are regions in the QGP depleted of energy and momentum due to the parton-medium scattering. Neither the recoil partons on the jet side nor the medium holes on the \PZ boson side rescatter with the medium.

The ``\hybrid model~\cite{Casalderrey-Solana:2014bpa}," based on the AdS/CFT correspondence, combines pQCD-based calculations with strong-coupling dynamics in order to simulate jet interactions within the QGP. This approach provides a comprehensive framework for understanding the energy loss and modifications of jets traversing the QGP. According to this model, a positive wake associated with the jet shower enhances the jet-side \zay, while a negative wake suppresses the yield around the \PZ bosons~\cite{Casalderrey-Solana:2020rsj}. 

Finally, the coupled linear Boltzmann transport (\colbt) model combines the linear Boltzmann transport equation, which governs parton energy loss through scattering and radiation in the QGP, with hydrodynamic simulations that describe the bulk evolution of the QGP~\cite{Chen:2017zte}.
In this model, ``reheating" refers to the process by which energy lost by high-energy partons is transferred back into the medium and becomes a source term in the hydrodynamic simulation, potentially increasing the local temperature and altering the QGP's properties~\cite{He:2015pra}. At low \ptt, the reheating is linked to the enhanced associated yield on the jet side of the \zay. A diffusion wake, the diffusion of the "particle holes" behind the hard-scattered parton, results in the suppression of the \zay on the \PZ boson side~\cite{Yang:2021qtl,Yang:2022nei}.

To summarize, for the implementation of medium response in these models, three approaches are used: no rescattering (\jewel), iterative recoil treatment with hydrodynamic updates (\colbt), and full thermalization of quenched energy and wake (\hybrid). These methods offer complementary insights into the medium's response to parton energy loss and provide an opportunity to examine the extent to which recoil partons are thermalized in the medium.

\begin{figure*}[ht!]
    \centering
    \includegraphics[width=0.8\textwidth]{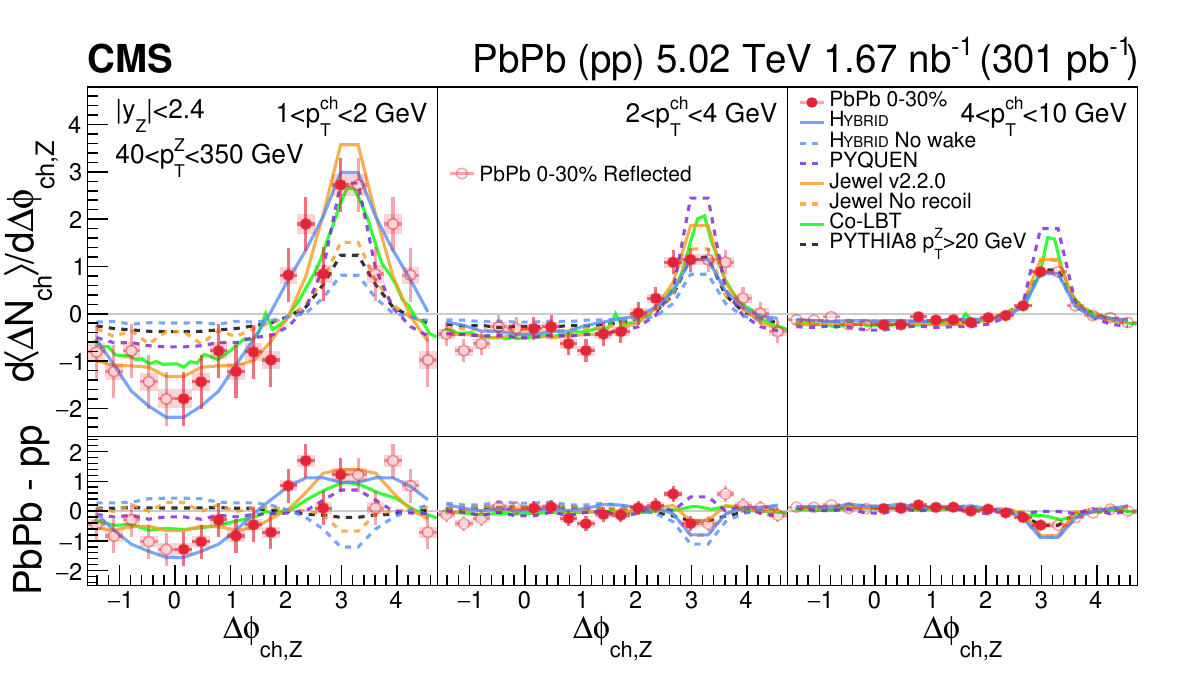}
    \caption{Upper row: Distributions of \dphitz in 0--30\% centrality \PbPb collisions in the charged hadron \ptt intervals of 1--2 (left column), 2--4 (middle column) and 4--10\GeV (right column). Lower row: The differences between the \PbPb and \pp distributions. The filled circles are \PbPb data and the open circles are reflected data. The vertical bars and shaded boxes represent the statistical and systematic uncertainties, respectively. The results are compared to predictions from theoretical models.}
    \label{fig:result-theoryDphi}
\end{figure*}

\begin{figure*}[ht!]
    \centering
    \includegraphics[width=0.8\textwidth]{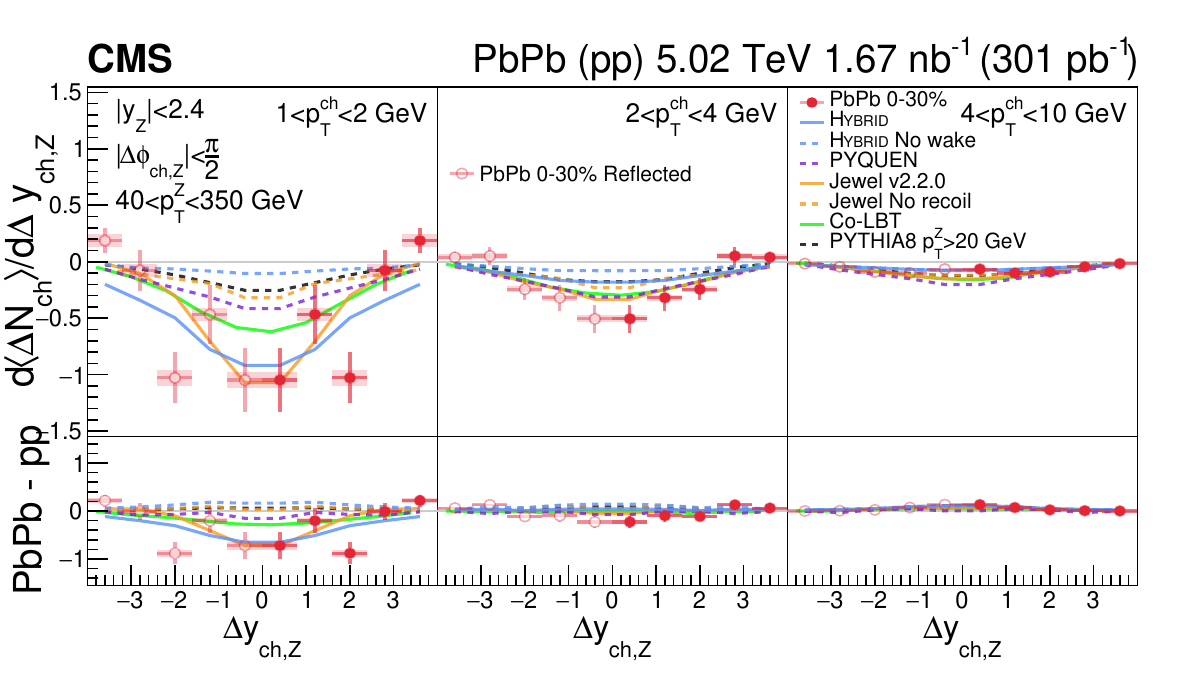}
    \caption{Upper row: Distributions of \dytz in 0--30\% centrality \PbPb collisions in the charged hadron \ptt intervals of 1--2 (left column), 2--4 (middle column) and 4--10\GeV (right column). Lower row: The differences between the \PbPb and \pp distributions. The filled circles are \PbPb data and the open circles are reflected data. The vertical bars and shaded boxes represent the statistical and systematic uncertainties, respectively. The results are compared to predictions from theoretical models.}
    \label{fig:result-theoryDeta}
\end{figure*}

Figures~\ref{fig:result-theoryDphi} and \ref{fig:result-theoryDeta} show comparisons between the 0--30\% centrality \PbPb data and theoretical predictions. The full \hybrid model predicts a dip structure at small $\dphitz$ and $\dytz$ due to the negative wake, and an excess at $\dphitz \sim \pi$ due to the positive wake. This model provides a good description of the central \PbPb data. However, if the wake contribution is excluded, the \hybrid model does not agree with data, particularly in the low-\pt region for $\dphitz\sim 0$ and $\dytz\sim 0$. The \jewel model correctly predicts the peak structure at $\dphitz = \pi$ as a result of medium recoils, which emphasizes their impact on the azimuthal distribution. Additionally, the dip in $\dytz$ on the \PZ boson side is also linked to medium recoils. Without medium recoils, \jewel fails to describe the data.  The \pyquen model predicts smaller modifications of the charged-hadron spectra and fails to describe the dip structures at small $\dphitz$ and $\dytz$ due to a lack of energy-momentum conservation. The \colbt model provides a reasonable description of the \dphitz spectra in the three ranges of \ptt. However, this model predicts a larger (i.e., less negative) \zay at small \dytz compared to the data in the  range $1 < \ptt < 4\GeV$ . Finally, the \PbPb data are compared to events simulated using
\PYTHIAviii with a selection of $\ptz > 20\GeV$, where the value is selected to roughly match the high-\ptt \zay distribution. This comparison is intended to determine if the higher energy quenched jets in the data, which have $\ptz > 40\GeV$, resemble lower energy jets in vacuum generated by the \PYTHIA~8 simulation. For charged particles with $1 < \ptt < 2\GeV$, the lower-\ptz~\PYTHIA predictions do not match the \PbPb data, indicating that quenched jets do not resemble lower-\pt vacuum-like jets.

Based on the comparisons of model predictions with the \PbPb data, a better description is found using theoretical models that include medium-recoil effects. In addition, a very poor description is found using simulations of lower-energy vacuum jets.

These results provide the first evidence of the presence of medium recoil and hole effects around the \PZ boson. However, greater statistical precision would be needed to determine which of the theoretical models including medium-recoil effects best matches the data.

\section{Summary}

This Letter presents the first measurement of correlations between \PZ bosons and charged hadrons in bins of charged-hadron transverse momentum (\pt) and event centrality for lead-lead (\PbPb) collisions at a nucleon-nucleon center of mass energy of $ 5.02\TeV$  and proton-proton (\pp) collisions at the same energy. By studying the hadrons produced along with the \PZ boson, the analysis probes parton-medium interactions, including modifications to parton showers and medium response. The charged-hadron rapidity and azimuthal angle distributions relative to the direction of \PZ bosons with $40 < \ptz < 350\GeV$  are extracted. The analysis utilizes pp (PbPb) data from 2017 (2018) with an integrated luminosity of $301$\pbinv ($1.67$\nbinv).

The \zay of charged hadrons in Z boson-tagged events is measured in bins of the differences in azimuthal angle ($\dphitz$) and in rapidity ($\dytz$) and is compared between the \PbPb and reference pp data. For low-\pt ($1 < \pt < 2\GeV$) charged hadrons in \PbPb collisions, a dip at $\dphitz \sim 0$ indicates the presence of a negative medium wake or medium holes. An excess at $\dphitz \sim \pi$ suggests medium-induced radiation, medium recoils, and/or momentum-broadening effects.
Central \PbPb collisions show larger modulation in the $\dphitz$ distribution than \pp data for these low-\pt charged hadrons, a difference which diminishes in more peripheral events.
For particles with higher \pt ( of $4-10\GeV$), a reduction in the jet-side \zay is consistent with jet quenching. The $\dytz$ distribution shows significant deviations in central \PbPb collisions with respect to the pp reference, especially at low charged-hadron \pt, where the \PbPb yield is lower than that in pp collisions near the \PZ boson ($\abs{\dytz} \sim 0$).

The \PbPb data are then compared to a variety of theoretical models including different implementations of jet energy loss and the medium response. Models that do not include medium response effects fail to describe the dip in the data at small $\dphitz$ and $\dytz$ expected from negative response contributions, and an excess at $\dphitz \sim \pi$ due to the positive response contributions. In comparison, the models including such response effects provide a better description of the \PbPb data. Amongst the different medium response implementations, for example that of a hydrodynamic positive and negative wake or medium recoils with corresponding hole contributions, it is not clear within the current statistical uncertainty which provides the best description of the data.

Overall, the differences between the \PbPb and \pp data, with a significance greater than $3\sigma$, provide the first evidence of medium-recoil and medium-hole effects caused by a hard probe since the \PbPb data are better described by theoretical models that include these effects. Comparisons with theoretical models highlight the unique insights these approaches offer into the properties of the QGP medium, especially its response to the energy loss experienced by partons which pass through it.

\begin{acknowledgments}
We congratulate our colleagues in the CERN accelerator departments for the excellent performance of the LHC and thank the technical and administrative staffs at CERN and at other CMS institutes for their contributions to the success of the CMS effort. In addition, we gratefully acknowledge the computing centers and personnel of the Worldwide LHC Computing Grid and other centers for delivering so effectively the computing infrastructure essential to our analyses. Finally, we acknowledge the enduring support for the construction and operation of the LHC, the CMS detector, and the supporting computing infrastructure provided by the following funding agencies: SC (Armenia), BMBWF and FWF (Austria); FNRS and FWO (Belgium); CNPq, CAPES, FAPERJ, FAPERGS, and FAPESP (Brazil); MES and BNSF (Bulgaria); CERN; CAS, MoST, and NSFC (China); MINCIENCIAS (Colombia); MSES and CSF (Croatia); RIF (Cyprus); SENESCYT (Ecuador); ERC PRG, RVTT3 and MoER TK202 (Estonia); Academy of Finland, MEC, and HIP (Finland); CEA and CNRS/IN2P3 (France); SRNSF (Georgia); BMBF, DFG, and HGF (Germany); GSRI (Greece); NKFIH (Hungary); DAE and DST (India); IPM (Iran); SFI (Ireland); INFN (Italy); MSIT and NRF (Republic of Korea); MES (Latvia); LMTLT (Lithuania); MOE and UM (Malaysia); BUAP, CINVESTAV, CONACYT, LNS, SEP, and UASLP-FAI (Mexico); MOS (Montenegro); MBIE (New Zealand); PAEC (Pakistan); MES and NSC (Poland); FCT (Portugal); MESTD (Serbia); MICIU/AEI and PCTI (Spain); MOSTR (Sri Lanka); Swiss Funding Agencies (Switzerland); MST (Taipei); MHESI and NSTDA (Thailand); TUBITAK and TENMAK (T\"{u}rkiye); NASU (Ukraine); STFC (United Kingdom); DOE and NSF (USA).

\hyphenation{Rachada-pisek} Individuals have received support from the Marie-Curie program and the European Research Council and Horizon 2020 Grant, contract Nos.\ 675440, 724704, 752730, 758316, 765710, 824093, 101115353, 101002207, 101001205, and COST Action CA16108 (European Union); the Leventis Foundation; the Alfred P.\ Sloan Foundation; the Alexander von Humboldt Foundation; the Science Committee, project no. 22rl-037 (Armenia); the Fonds pour la Formation \`a la Recherche dans l'Industrie et dans l'Agriculture (FRIA-Belgium); the Beijing Municipal Science \& Technology Commission, No. Z191100007219010, the Fundamental Research Funds for the Central Universities, the Ministry of Science and Technology of China under Grant No. 2023YFA1605804, and the Natural Science Foundation of China under Grant No. 12061141002 (China); the Ministry of Education, Youth and Sports (MEYS) of the Czech Republic; the Shota Rustaveli National Science Foundation, grant FR-22-985 (Georgia); the Deutsche Forschungsgemeinschaft (DFG), among others, under Germany's Excellence Strategy -- EXC 2121 ``Quantum Universe" -- 390833306, and under project number 400140256 - GRK2497; the Hellenic Foundation for Research and Innovation (HFRI), Project Number 2288 (Greece); the Hungarian Academy of Sciences, the New National Excellence Program - \'UNKP, the NKFIH research grants K 131991, K 133046, K 138136, K 143460, K 143477, K 146913, K 146914, K 147048, 2020-2.2.1-ED-2021-00181, TKP2021-NKTA-64, and 2021-4.1.2-NEMZ\_KI-2024-00036 (Hungary); the Council of Science and Industrial Research, India; ICSC -- National Research Center for High Performance Computing, Big Data and Quantum Computing, FAIR -- Future Artificial Intelligence Research, and CUP I53D23001070006 (Mission 4 Component 1), funded by the NextGenerationEU program (Italy); the Latvian Council of Science; the Ministry of Education and Science, project no. 2022/WK/14, and the National Science Center, contracts Opus 2021/41/B/ST2/01369 and 2021/43/B/ST2/01552 (Poland); the Funda\c{c}\~ao para a Ci\^encia e a Tecnologia, grant CEECIND/01334/2018 (Portugal); the National Priorities Research Program by Qatar National Research Fund; MICIU/AEI/10.13039/501100011033, ERDF/EU, "European Union NextGenerationEU/PRTR", and Programa Severo Ochoa del Principado de Asturias (Spain); the Chulalongkorn Academic into Its 2nd Century Project Advancement Project, and the National Science, Research and Innovation Fund via the Program Management Unit for Human Resources \& Institutional Development, Research and Innovation, grant B39G680009 (Thailand); the Kavli Foundation; the Nvidia Corporation; the SuperMicro Corporation; the Welch Foundation, contract C-1845; and the Weston Havens Foundation (USA).
\end{acknowledgments}\section*{Data availability} Release and preservation of data used by the CMS Collaboration as the basis for publications is guided by the  \href{https://doi.org/10.7483/OPENDATA.CMS.1BNU.8V1W}{CMS data preservation, re-use and open access policy}.

\bibliography{auto_generated}

\providecommand{\href}[2]{#2}\begingroup\raggedright\begin{thebibliography}{10}%
\makeatletter
\providecommand{\hrefCMSnoop }[0]{\@secondoftwo}%
\makeatother
\providecommand{\doi}{\texttt{doi:}\begingroup \urlstyle{tt}\Url}

\bibitem{PhysRevLett.34.1353}
\hrefCMSnoop {}{J.~C. Collins and M.~J. Perry, ``Superdense matter: Neutrons or
  asymptotically free quarks?'',} \textit{ Phys. Rev. Lett.} \textbf{ 34}
  (1975) 1353,
  \href{http://dx.doi.org/10.1103/PhysRevLett.34.1353}{\doi{10.1103/PhysRevLett.34.1353}}.

\bibitem{Karsch:1995sy}
\hrefCMSnoop {}{F.~Karsch, ``{The phase transition to the quark gluon plasma:
  recent results from lattice calculations}'',} \textit{ Nucl. Phys. A}
  \textbf{ 590} (1995) 367,
  \href{http://dx.doi.org/10.1016/0375-9474(95)00248-Y}{\doi{10.1016/0375-9474(95)00248-Y}},
\href{http://www.arXiv.org/abs/hep-lat/9503010}{\texttt{arXiv:hep-lat/9503010}}.

\bibitem{Busza:2018rrf}
\hrefCMSnoop {}{W.~Busza, K.~Rajagopal, and W.~van~der Schee, ``{Heavy ion
  Collisions: The big picture, and the big questions}'',} \textit{ Ann. Rev.
  Nucl. Part. Sci.} \textbf{ 68} (2018) 339,
  \href{http://dx.doi.org/10.1146/annurev-nucl-101917-020852}{\doi{10.1146/annurev-nucl-101917-020852}},
  \href{http://www.arXiv.org/abs/1802.04801}{\texttt{arXiv:1802.04801}}.

\bibitem{STAR:2005gfr}
\hrefCMSnoop {}{{STAR} Collaboration, ``{Experimental and theoretical
  challenges in the search for the quark gluon plasma: The STAR Collaboration's
  critical assessment of the evidence from RHIC collisions}'',} \textit{ Nucl.
  Phys. A} \textbf{ 757} (2005) 102,
  \href{http://dx.doi.org/10.1016/j.nuclphysa.2005.03.085}{\doi{10.1016/j.nuclphysa.2005.03.085}},
  \href{http://www.arXiv.org/abs/nucl-ex/0501009}{\texttt{arXiv:nucl-ex/0501009}}.

\bibitem{PHENIX:2004vcz}
\hrefCMSnoop {}{{PHENIX} Collaboration, ``{Formation of dense partonic matter
  in relativistic nucleus-nucleus collisions at RHIC: Experimental evaluation
  by the PHENIX collaboration}'',} \textit{ Nucl. Phys. A} \textbf{ 757} (2005)
  184,
  \href{http://dx.doi.org/10.1016/j.nuclphysa.2005.03.086}{\doi{10.1016/j.nuclphysa.2005.03.086}},
  \href{http://www.arXiv.org/abs/nucl-ex/0410003}{\texttt{arXiv:nucl-ex/0410003}}.

\bibitem{PHOBOS:2004zne}
\hrefCMSnoop {}{{PHOBOS} Collaboration, ``{The PHOBOS perspective on
  discoveries at RHIC}'',} \textit{ Nucl. Phys. A} \textbf{ 757} (2005) 28,
  \href{http://dx.doi.org/10.1016/j.nuclphysa.2005.03.084}{\doi{10.1016/j.nuclphysa.2005.03.084}},
  \href{http://www.arXiv.org/abs/nucl-ex/0410022}{\texttt{arXiv:nucl-ex/0410022}}.

\bibitem{BRAHMS:2004adc}
\hrefCMSnoop {}{{BRAHMS} Collaboration, ``{Quark gluon plasma and color glass
  condensate at RHIC? The perspective from the BRAHMS experiment}'',} \textit{
  Nucl. Phys. A} \textbf{ 757} (2005) 1,
  \href{http://dx.doi.org/10.1016/j.nuclphysa.2005.02.130}{\doi{10.1016/j.nuclphysa.2005.02.130}},
  \href{http://www.arXiv.org/abs/nucl-ex/0410020}{\texttt{arXiv:nucl-ex/0410020}}.

\bibitem{ALICE:2022wpn}
\hrefCMSnoop {}{{ALICE Collaboration}, ``{The ALICE experiment: a journey
  through QCD}'',} \textit{ Eur. Phys. J. C} \textbf{ 84} (2024) 813,
  \href{http://dx.doi.org/10.1140/epjc/s10052-024-12935-y}{\doi{10.1140/epjc/s10052-024-12935-y}},
  \href{http://www.arXiv.org/abs/2211.04384}{\texttt{arXiv:2211.04384}}.

\bibitem{CMS:2024krd}
\hrefCMSnoop {}{{CMS Collaboration}, ``{Overview of high-density QCD studies
  with the CMS experiment at the LHC}'',} \textit{ Physics Reports} \textbf{
  1115} (2025) 219,
  \href{http://dx.doi.org/10.1016/j.physrep.2024.11.007}{\doi{10.1016/j.physrep.2024.11.007}},
  \href{http://www.arXiv.org/abs/2405.10785}{\texttt{arXiv:2405.10785}}.

\bibitem{PhysRevD.27.140}
\hrefCMSnoop {}{J.~D. Bjorken, ``Highly relativistic nucleus-nucleus
  collisions: The central rapidity region'',} \textit{ Phys. Rev. D} \textbf{
  27} (1983) 140,
  \href{http://dx.doi.org/10.1103/PhysRevD.27.140}{\doi{10.1103/PhysRevD.27.140}}.

\bibitem{Gyulassy:1990ye}
\hrefCMSnoop {}{M.~Gyulassy and M.~Pl{\"u}mer, ``{Jet quenching in dense
  matter}'',} \textit{ Phys. Lett. B} \textbf{ 243} (1990) 432,
\href{http://dx.doi.org/10.1016/0370-2693(90)91409-5}{\doi{10.1016/0370-2693(90)91409-5}}.

\bibitem{Adams:2003kv}
\hrefCMSnoop {}{{STAR} Collaboration, ``Transverse-momentum and
  collision-energy dependence of high-$p_{\mathrm{t}}$ hadron suppression in
  {Au+Au} collisions at ultrarelativistic energies'',} \textit{ Phys. Rev.
  Lett.} \textbf{ 91} (2003) 172302,
  \href{http://dx.doi.org/10.1103/PhysRevLett.91.172302}{\doi{10.1103/PhysRevLett.91.172302}},
\href{http://www.arXiv.org/abs/nucl-ex/0305015}{\texttt{arXiv:nucl-ex/0305015}}.

\bibitem{Adare:2008qa}
\hrefCMSnoop {}{{PHENIX} Collaboration, ``Suppression pattern of neutral pions
  at high transverse momentum in {Au+Au} collisions at {$\rootsNN = 200$\GeV}
  and constraints on medium transport coefficients'',} \textit{ Phys. Rev.
  Lett.} \textbf{ 101} (2008) 232301,
  \href{http://dx.doi.org/10.1103/PhysRevLett.101.232301}{\doi{10.1103/PhysRevLett.101.232301}},
\href{http://www.arXiv.org/abs/0801.4020}{\texttt{arXiv:0801.4020}}.

\bibitem{Abelev:2012hxa}
\hrefCMSnoop {}{{ALICE Collaboration}, ``{Centrality dependence of charged
  particle production at large transverse momentum in Pb--Pb collisions at
  $\rootsNN = 2.76$\TeV}'',} \textit{ Phys. Lett. B} \textbf{ 720} (2013) 52,
  \href{http://dx.doi.org/10.1016/j.physletb.2013.01.051}{\doi{10.1016/j.physletb.2013.01.051}},
\href{http://www.arXiv.org/abs/1208.2711}{\texttt{arXiv:1208.2711}}.

\bibitem{Aad:2015wga}
\hrefCMSnoop {}{{ATLAS Collaboration}, ``{Measurement of charged-particle
  spectra in Pb+Pb collisions at $\rootsNN = 2.76\TeV$ with the ATLAS detector
  at the LHC}'',} \textit{ JHEP} \textbf{ 09} (2015) 050,
  \href{http://dx.doi.org/10.1007/JHEP09(2015)050}{\doi{10.1007/JHEP09(2015)050}},
\href{http://www.arXiv.org/abs/1504.04337}{\texttt{arXiv:1504.04337}}.

\bibitem{CMS:2012aa}
\hrefCMSnoop {}{{CMS Collaboration}, ``{Study of high-$p_{\rm{T}}$ charged
  particle suppression in PbPb compared to pp collisions at
  $\rootsNN=2.76$\TeV}'',} \textit{ Eur. Phys. J. C} \textbf{ 72} (2012) 1945,
  \href{http://dx.doi.org/10.1140/epjc/s10052-012-1945-x}{\doi{10.1140/epjc/s10052-012-1945-x}},
\href{http://www.arXiv.org/abs/1202.2554}{\texttt{arXiv:1202.2554}}.

\bibitem{Khachatryan:2016odn}
\hrefCMSnoop {}{{CMS Collaboration}, ``{Charged-particle nuclear modification
  factors in PbPb and pPb collisions at $\rootsNN=5.02\TeV$}'',} \textit{ JHEP}
  \textbf{ 04} (2017) 039,
  \href{http://dx.doi.org/10.1007/JHEP04(2017)039}{\doi{10.1007/JHEP04(2017)039}},
\href{http://www.arXiv.org/abs/1611.01664}{\texttt{arXiv:1611.01664}}.

\bibitem{Chatrchyan:2011sx}
\hrefCMSnoop {}{{CMS Collaboration}, ``{Observation and studies of jet
  quenching in PbPb collisions at $\rootsNN = 2.76\TeV$}'',} \textit{ Phys.
  Rev. C} \textbf{ 84} (2011) 024906,
  \href{http://dx.doi.org/10.1103/PhysRevC.84.024906}{\doi{10.1103/PhysRevC.84.024906}},
\href{http://www.arXiv.org/abs/1102.1957}{\texttt{arXiv:1102.1957}}.

\bibitem{Aad:2010bu}
\hrefCMSnoop {}{{ATLAS Collaboration}, ``{Observation of a centrality-dependent
  dijet asymmetry in lead-lead collisions at $\rootsNN=2.76\TeV$ with the ATLAS
  detector at the LHC}'',} \textit{ Phys. Rev. Lett.} \textbf{ 105} (2010)
  252303,
  \href{http://dx.doi.org/10.1103/PhysRevLett.105.252303}{\doi{10.1103/PhysRevLett.105.252303}},
\href{http://www.arXiv.org/abs/1011.6182}{\texttt{arXiv:1011.6182}}.

\bibitem{ATLAS:2014cpa}
\hrefCMSnoop {}{{ATLAS Collaboration}, ``{Centrality and rapidity dependence of
  inclusive jet production in $\rootsNN = 5.02$\TeV proton--lead collisions
  with the ATLAS detector}'',} \textit{ Phys. Lett. B} \textbf{ 748} (2015)
  392,
  \href{http://dx.doi.org/10.1016/j.physletb.2015.07.023}{\doi{10.1016/j.physletb.2015.07.023}},
\href{http://www.arXiv.org/abs/1412.4092}{\texttt{arXiv:1412.4092}}.

\bibitem{Adam:2015doa}
\hrefCMSnoop {}{{ALICE Collaboration}, ``{Measurement of jet quenching with
  semi-inclusive hadron-jet distributions in central Pb-Pb collisions at
  $\rootsNN=2.76\TeV$}'',} \textit{ JHEP} \textbf{ 09} (2015) 170,
  \href{http://dx.doi.org/10.1007/JHEP09(2015)170}{\doi{10.1007/JHEP09(2015)170}},
\href{http://www.arXiv.org/abs/1506.03984}{\texttt{arXiv:1506.03984}}.

\bibitem{Khachatryan:2016jfl}
\hrefCMSnoop {}{{CMS Collaboration}, ``{Measurement of inclusive jet cross
  sections in pp and PbPb collisions at $\rootsNN=2.76\TeV$}'',} \textit{ Phys.
  Rev. C} \textbf{ 96} (2017) 015202,
  \href{http://dx.doi.org/10.1103/PhysRevC.96.015202}{\doi{10.1103/PhysRevC.96.015202}},
\href{http://www.arXiv.org/abs/1609.05383}{\texttt{arXiv:1609.05383}}.

\bibitem{Adam:2015ewa}
\hrefCMSnoop {}{{ALICE Collaboration}, ``{Measurement of jet suppression in
  central Pb-Pb collisions at $\rootsNN = 2.76\TeV$}'',} \textit{ Phys. Lett.
  B} \textbf{ 746} (2015) 1,
  \href{http://dx.doi.org/10.1016/j.physletb.2015.04.039}{\doi{10.1016/j.physletb.2015.04.039}},
\href{http://www.arXiv.org/abs/1502.01689}{\texttt{arXiv:1502.01689}}.

\bibitem{Adamczyk:2016fqm}
\hrefCMSnoop {}{{STAR} Collaboration, ``{Dijet imbalance measurements in Au+Au
  and pp collisions at $\sqrt{s_{_{\mathrm{NN}}}} = 200\GeV$ at STAR}'',}
  \textit{ Phys. Rev. Lett.} \textbf{ 119} (2017) 062301,
  \href{http://dx.doi.org/10.1103/PhysRevLett.119.062301}{\doi{10.1103/PhysRevLett.119.062301}},
\href{http://www.arXiv.org/abs/1609.03878}{\texttt{arXiv:1609.03878}}.

\bibitem{Casalderrey-Solana:2014bpa}
J.~Casalderrey-Solana\hrefCMSnoop {}{ { et~al.}, ``{A hybrid strong/weak
  coupling approach to jet quenching}'',} \textit{ JHEP} \textbf{ 10} (2014)
  019,
  \href{http://dx.doi.org/10.1007/JHEP10(2014)019}{\doi{10.1007/JHEP10(2014)019}},
  \href{http://www.arXiv.org/abs/1405.3864}{\texttt{arXiv:1405.3864}}.
Erratum: \DOI{10.1007/JHEP09(2015)175}.

\bibitem{Cao:2020wlm}
\hrefCMSnoop {}{S.~Cao and X.-N. Wang, ``{Jet quenching and medium response in
  high-energy heavy-ion collisions: a review}'',} \textit{ Rept. Prog. Phys.}
  \textbf{ 84} (2021) 024301,
  \href{http://dx.doi.org/10.1088/1361-6633/abc22b}{\doi{10.1088/1361-6633/abc22b}},
  \href{http://www.arXiv.org/abs/2002.04028}{\texttt{arXiv:2002.04028}}.

\bibitem{Apolinario:2022vzg}
\hrefCMSnoop {}{L.~Apolin{\'ario}, Y.-J. Lee, and M.~Winn, ``{Heavy quarks and
  jets as probes of the QGP}'',} \textit{ Prog. Part. Nucl. Phys.} \textbf{
  127} (2022) 103990,
  \href{http://dx.doi.org/10.1016/j.ppnp.2022.103990}{\doi{10.1016/j.ppnp.2022.103990}},
  \href{http://www.arXiv.org/abs/2203.16352}{\texttt{arXiv:2203.16352}}.

\bibitem{Cao:2022odi}
\hrefCMSnoop {}{S.~Cao and G.-Y. Qin, ``{Medium response and
  jet\textendash{}hadron correlations in relativistic heavy-ion collisions}'',}
  \textit{ Ann. Rev. Nucl. Part. Sci.} \textbf{ 73} (2023) 205,
  \href{http://dx.doi.org/10.1146/annurev-nucl-112822-031317}{\doi{10.1146/annurev-nucl-112822-031317}},
  \href{http://www.arXiv.org/abs/2211.16821}{\texttt{arXiv:2211.16821}}.

\bibitem{Chatrchyan:2012vq}
\hrefCMSnoop {}{{CMS Collaboration}, ``{Measurement of isolated photon
  production in pp and PbPb collisions at $\rootsNN = 2.76\TeV$}'',} \textit{
  Phys. Lett. B} \textbf{ 710} (2012) 256,
  \href{http://dx.doi.org/10.1016/j.physletb.2012.02.077}{\doi{10.1016/j.physletb.2012.02.077}},
\href{http://www.arXiv.org/abs/1201.3093}{\texttt{arXiv:1201.3093}}.

\bibitem{Chatrchyan:2012nt}
\hrefCMSnoop {}{{CMS Collaboration}, ``{Study of W boson production in PbPb and
  pp collisions at $\rootsNN=2.76$\TeV}'',} \textit{ Phys. Lett. B} \textbf{
  715} (2012) 66,
  \href{http://dx.doi.org/10.1016/j.physletb.2012.07.025}{\doi{10.1016/j.physletb.2012.07.025}},
\href{http://www.arXiv.org/abs/1205.6334}{\texttt{arXiv:1205.6334}}.

\bibitem{Chatrchyan:2014csa}
\hrefCMSnoop {}{{CMS Collaboration}, ``{Study of Z production in PbPb and pp
  collisions at $ \rootsNN =2.76$\TeV in the dimuon and dielectron decay
  channels}'',} \textit{ JHEP} \textbf{ 03} (2015) 022,
  \href{http://dx.doi.org/10.1007/JHEP03(2015)022}{\doi{10.1007/JHEP03(2015)022}},
\href{http://www.arXiv.org/abs/1410.4825}{\texttt{arXiv:1410.4825}}.

\bibitem{ATLAS:2015rlt}
\hrefCMSnoop {}{{ATLAS Collaboration}, ``{Centrality, rapidity and transverse
  momentum dependence of isolated prompt photon production in lead-lead
  collisions at $\rootsNN = 2.76$ TeV measured with the ATLAS detector}'',}
  \textit{ Phys. Rev. C} \textbf{ 93} (2016) 034914,
  \href{http://dx.doi.org/10.1103/PhysRevC.93.034914}{\doi{10.1103/PhysRevC.93.034914}},
  \href{http://www.arXiv.org/abs/1506.08552}{\texttt{arXiv:1506.08552}}.

\bibitem{CMS:2020oen}
\hrefCMSnoop {}{{CMS Collaboration}, ``{The production of isolated photons in
  PbPb and pp collisions at $\rootsNN =$ 5.02 TeV}'',} \textit{ JHEP} \textbf{
  07} (2020) 116,
  \href{http://dx.doi.org/10.1007/JHEP07(2020)116}{\doi{10.1007/JHEP07(2020)116}},
  \href{http://www.arXiv.org/abs/2003.12797}{\texttt{arXiv:2003.12797}}.

\bibitem{CMS:2021kvd}
\hrefCMSnoop {}{{CMS Collaboration}, ``{Constraints on the initial state of
  Pb-Pb collisions via measurements of $Z$-boson yields and azimuthal
  anisotropy at $\rootsNN$ = 5.02\TeV}'',} \textit{ Phys. Rev. Lett.} \textbf{
  127} (2021) 102002,
  \href{http://dx.doi.org/10.1103/PhysRevLett.127.102002}{\doi{10.1103/PhysRevLett.127.102002}},
  \href{http://www.arXiv.org/abs/2103.14089}{\texttt{arXiv:2103.14089}}.

\bibitem{Kartvelishvili:1995fr}
\hrefCMSnoop {}{V.~Kartvelishvili, R.~Kvatadze, and R.~Shanidze, ``{On Z and
  Z+jet production in heavy ion collisions}'',} \textit{ Phys. Lett. B}
  \textbf{ 356} (1995) 589,
  \href{http://dx.doi.org/10.1016/0370-2693(95)00865-I}{\doi{10.1016/0370-2693(95)00865-I}},
\href{http://www.arXiv.org/abs/hep-ph/9505418}{\texttt{arXiv:hep-ph/9505418}}.

\bibitem{Wang:1996yh}
\hrefCMSnoop {}{X.-N. Wang, Z.~Huang, and I.~Sarcevic, ``{Jet quenching in the
  opposite direction of a tagged photon in high-energy heavy ion
  collisions}'',} \textit{ Phys. Rev. Lett.} \textbf{ 77} (1996) 231,
  \href{http://dx.doi.org/10.1103/PhysRevLett.77.231}{\doi{10.1103/PhysRevLett.77.231}},
\href{http://www.arXiv.org/abs/hep-ph/9605213}{\texttt{arXiv:hep-ph/9605213}}.

\bibitem{Wang:1996pe}
\hrefCMSnoop {}{X.-N. Wang and Z.~Huang, ``{Medium-induced parton energy loss
  in $\ensuremath{\gamma}$+jet events of high-energy heavy-ion collisions}'',}
  \textit{ Phys. Rev. C} \textbf{ 55} (1997) 3047,
  \href{http://dx.doi.org/10.1103/PhysRevC.55.3047}{\doi{10.1103/PhysRevC.55.3047}},
\href{http://www.arXiv.org/abs/hep-ph/9701227}{\texttt{arXiv:hep-ph/9701227}}.

\bibitem{Brewer:2021hmh}
\hrefCMSnoop {}{J.~Brewer, Q.~Brodsky, and K.~Rajagopal, ``{Disentangling jet
  modification in jet simulations and in Z+jet data}'',} \textit{ JHEP}
  \textbf{ 02} (2022) 175,
  \href{http://dx.doi.org/10.1007/JHEP02(2022)175}{\doi{10.1007/JHEP02(2022)175}},
  \href{http://www.arXiv.org/abs/2110.13159}{\texttt{arXiv:2110.13159}}.

\bibitem{Dai:2012am}
\hrefCMSnoop {}{W.~Dai, I.~Vitev, and B.-W. Zhang, ``{Momentum imbalance of
  isolated photon-tagged jet production at RHIC and LHC}'',} \textit{ Phys.
  Rev. Lett.} \textbf{ 110} (2013) 142001,
  \href{http://dx.doi.org/10.1103/PhysRevLett.110.142001}{\doi{10.1103/PhysRevLett.110.142001}},
\href{http://www.arXiv.org/abs/1207.5177}{\texttt{arXiv:1207.5177}}.

\bibitem{Kang:2017xnc}
\hrefCMSnoop {}{Z.-B. Kang, I.~Vitev, and H.~Xing, ``{Vector-boson-tagged jet
  production in heavy ion collisions at energies available at the CERN large
  hadron collider}'',} \textit{ Phys. Rev. C} \textbf{ 96} (2017) 014912,
  \href{http://dx.doi.org/10.1103/PhysRevC.96.014912}{\doi{10.1103/PhysRevC.96.014912}},
\href{http://www.arXiv.org/abs/1702.07276}{\texttt{arXiv:1702.07276}}.

\bibitem{Sirunyan:2018qec}
\hrefCMSnoop {}{{CMS Collaboration}, ``{Observation of medium-induced
  modifications of jet fragmentation in Pb-Pb collisions at $\rootsNN=5.02\TeV$
  using isolated photon-tagged jets}'',} \textit{ Phys. Rev. Lett.} \textbf{
  121} (2018) 242301,
  \href{http://dx.doi.org/10.1103/PhysRevLett.121.242301}{\doi{10.1103/PhysRevLett.121.242301}},
\href{http://www.arXiv.org/abs/1801.04895}{\texttt{arXiv:1801.04895}}.

\bibitem{Sirunyan:2018ncy}
\hrefCMSnoop {}{{CMS Collaboration}, ``{Jet shapes of isolated photon-tagged
  jets in Pb-Pb and pp collisions at $\rootsNN=5.02\TeV$}'',} \textit{ Phys.
  Rev. Lett.} \textbf{ 122} (2019) 152001,
  \href{http://dx.doi.org/10.1103/PhysRevLett.122.152001}{\doi{10.1103/PhysRevLett.122.152001}},
\href{http://www.arXiv.org/abs/1809.08602}{\texttt{arXiv:1809.08602}}.

\bibitem{ATLAS:2024prm}
\hrefCMSnoop {}{{ATLAS Collaboration}, ``Search for the jet-induced diffusion
  wake in the quark-gluon plasma via measurements of jet-track correlations in
  photon-jet events in {Pb+Pb} collisions at {$\rootsNN=5.02\TeV$} with the
  {ATLAS} detector'',} \textit{ Phys. Rev. C} \textbf{ 111} (2025) 044909,
  \href{http://dx.doi.org/10.1103/PhysRevC.111.044909}{\doi{10.1103/PhysRevC.111.044909}},
  \href{http://www.arXiv.org/abs/2408.08599}{\texttt{arXiv:2408.08599}}.

\bibitem{CMS:2017eqd}
\hrefCMSnoop {}{{CMS Collaboration}, ``{Study of jet quenching with $Z+$jet
  correlations in Pb-Pb and pp collisions at $\rootsNN=5.02\text{ }\text{
  }\mathrm{TeV}$}'',} \textit{ Phys. Rev. Lett.} \textbf{ 119} (2017) 082301,
  \href{http://dx.doi.org/10.1103/PhysRevLett.119.082301}{\doi{10.1103/PhysRevLett.119.082301}},
  \href{http://www.arXiv.org/abs/1702.01060}{\texttt{arXiv:1702.01060}}.

\bibitem{CMS:2021otx}
\hrefCMSnoop {}{{CMS Collaboration}, ``{Using Z boson events to study
  parton-medium interactions in Pb-Pb collisions}'',} \textit{ Phys. Rev.
  Lett.} \textbf{ 128} (2022) 122301,
  \href{http://dx.doi.org/10.1103/PhysRevLett.128.122301}{\doi{10.1103/PhysRevLett.128.122301}},
  \href{http://www.arXiv.org/abs/2103.04377}{\texttt{arXiv:2103.04377}}.

\bibitem{Yang:2021qtl}
Z.~Yang\hrefCMSnoop {}{ { et~al.}, ``{Search for the elusive jet-induced
  diffusion wake in $Z/\gamma$-jets with 2D jet tomography in high-energy
  heavy-ion collisions}'',} \textit{ Phys. Rev. Lett.} \textbf{ 127} (2021)
  082301,
  \href{http://dx.doi.org/10.1103/PhysRevLett.127.082301}{\doi{10.1103/PhysRevLett.127.082301}},
  \href{http://www.arXiv.org/abs/2101.05422}{\texttt{arXiv:2101.05422}}.

\bibitem{Yang:2022nei}
Z.~Yang\hrefCMSnoop {}{ { et~al.}, ``{3D Structure of jet-induced diffusion
  wake in an expanding quark-gluon plasma}'',} \textit{ Phys. Rev. Lett.}
  \textbf{ 130} (2023) 052301,
  \href{http://dx.doi.org/10.1103/PhysRevLett.130.052301}{\doi{10.1103/PhysRevLett.130.052301}},
  \href{http://www.arXiv.org/abs/2203.03683}{\texttt{arXiv:2203.03683}}.

\bibitem{ATLAS:2020wmg}
\hrefCMSnoop {}{{ATLAS Collaboration}, ``Medium-induced modification of
  {Z}-tagged charged particle yields in {Pb+Pb} collisions at 5.02 {TeV} with
  the {ATLAS} detector'',} \textit{ Phys. Rev. Lett.} \textbf{ 126} (2021)
  072301,
  \href{http://dx.doi.org/10.1103/PhysRevLett.126.072301}{\doi{10.1103/PhysRevLett.126.072301}},
  \href{http://www.arXiv.org/abs/2008.09811}{\texttt{arXiv:2008.09811}}.

\bibitem{DEramo:2012uzl}
\hrefCMSnoop {}{F.~D'Eramo, M.~Lekaveckas, H.~Liu, and K.~Rajagopal,
  ``{Momentum broadening in weakly coupled quark-gluon plasma (with a view to
  finding the quasiparticles within liquid quark-gluon plasma)}'',} \textit{
  JHEP} \textbf{ 05} (2013) 031,
  \href{http://dx.doi.org/10.1007/JHEP05(2013)031}{\doi{10.1007/JHEP05(2013)031}},
  \href{http://www.arXiv.org/abs/1211.1922}{\texttt{arXiv:1211.1922}}.

\bibitem{He:2015pra}
\hrefCMSnoop {}{Y.~He, T.~Luo, X.-N. Wang, and Y.~Zhu, ``{Linear Boltzmann
  transport for jet propagation in the quark-gluon plasma: Elastic processes
  and medium recoil}'',} \textit{ Phys. Rev. C} \textbf{ 91} (2015) 054908,
  \href{http://dx.doi.org/10.1103/PhysRevC.91.054908}{\doi{10.1103/PhysRevC.91.054908}},
  \href{http://www.arXiv.org/abs/1503.03313}{\texttt{arXiv:1503.03313}}.
  [Erratum: $Phys.Rev.C$ $\textbf{97}$, 019902 (2018)].

\bibitem{CMS:2025rzq}
\hrefCMSnoop {}{{CMS Collaboration}, ``{Luminosity measurement for lead-lead
  collisions at $\sqrt{s_{\mathrm{NN}}}$ = 5.02 TeV in 2015 and 2018 at
  CMS}'',} 2025.
  \href{http://www.arXiv.org/abs/2503.03946}{\texttt{arXiv:2503.03946}}.
  Submitted to {EPJC}.

\bibitem{hepdata}
\hrefCMSnoop {}{{CMS Collaboration}, ``{HEPD}ata record for this analysis'',}
  2025.
\newblock
  \href{http://dx.doi.org/10.17182/hepdata.156186}{\doi{10.17182/hepdata.156186}}.

\bibitem{Chatrchyan:2008zzk}
\hrefCMSnoop {}{{CMS Collaboration}, ``The {CMS} experiment at the {CERN}
  {LHC}'',} \textit{ JINST} \textbf{ 3} (2008) S08004,
\href{http://dx.doi.org/10.1088/1748-0221/3/08/S08004}{\doi{10.1088/1748-0221/3/08/S08004}}.

\bibitem{Alwall:2014hca}
J.~Alwall\hrefCMSnoop {}{ { et~al.}, ``{The automated computation of tree-level
  and next-to-leading order differential cross sections, and their matching to
  parton shower simulations}'',} \textit{ JHEP} \textbf{ 07} (2014) 079,
  \href{http://dx.doi.org/10.1007/JHEP07(2014)079}{\doi{10.1007/JHEP07(2014)079}},
\href{http://www.arXiv.org/abs/1405.0301}{\texttt{arXiv:1405.0301}}.

\bibitem{Sjostrand:2014zea}
T.~Sj{\"o}strand\hrefCMSnoop {}{ { et~al.}, ``{An Introduction to PYTHIA
  8.2}'',} \textit{ Comput. Phys. Commun.} \textbf{ 191} (2015) 159,
  \href{http://dx.doi.org/10.1016/j.cpc.2015.01.024}{\doi{10.1016/j.cpc.2015.01.024}},
\href{http://www.arXiv.org/abs/1410.3012}{\texttt{arXiv:1410.3012}}.

\bibitem{CMS:2019csb}
\hrefCMSnoop {}{{CMS Collaboration}, ``{Extraction and validation of a new set
  of CMS PYTHIA8 tunes from underlying-event measurements}'',} \textit{ Eur.
  Phys. J. C} \textbf{ 80} (2020) 4,
  \href{http://dx.doi.org/10.1140/epjc/s10052-019-7499-4}{\doi{10.1140/epjc/s10052-019-7499-4}},
  \href{http://www.arXiv.org/abs/1903.12179}{\texttt{arXiv:1903.12179}}.

\bibitem{Lokhtin:2005px}
\hrefCMSnoop {}{I.~P. Lokhtin and A.~M. Snigirev, ``{A model of jet quenching
  in ultrarelativistic heavy ion collisions and high-$p_{\mathrm{T}}$ hadron
  spectra at RHIC}'',} \textit{ Eur. Phys. J. C} \textbf{ 45} (2006) 211,
  \href{http://dx.doi.org/10.1140/epjc/s2005-02426-3}{\doi{10.1140/epjc/s2005-02426-3}},
\href{http://www.arXiv.org/abs/hep-ph/0506189}{\texttt{arXiv:hep-ph/0506189}}.

\bibitem{geant4}
\hrefCMSnoop {}{{GEANT4} Collaboration, ``{\GEANTfour}---a simulation
  toolkit'',} \textit{ Nucl. Instrum. Meth. A} \textbf{ 506} (2003) 250,
  \href{http://dx.doi.org/10.1016/S0168-9002(03)01368-8}{\doi{10.1016/S0168-9002(03)01368-8}}.

\bibitem{Khachatryan:2016bia}
\hrefCMSnoop {}{{CMS Collaboration}, ``The {CMS} trigger system'',} \textit{
  JINST} \textbf{ 12} (2017) P01020,
  \href{http://dx.doi.org/10.1088/1748-0221/12/01/P01020}{\doi{10.1088/1748-0221/12/01/P01020}},
\href{http://www.arXiv.org/abs/1609.02366}{\texttt{arXiv:1609.02366}}.

\bibitem{CMS:2024qgs}
\hrefCMSnoop {}{{CMS Collaboration}, ``{Performance of CMS muon reconstruction
  from proton-proton to heavy ion collisions}'',} \textit{ JINST} \textbf{ 19}
  (2024) P09012,
  \href{http://dx.doi.org/10.1088/1748-0221/19/09/P09012}{\doi{10.1088/1748-0221/19/09/P09012}},
  \href{http://www.arXiv.org/abs/2404.17377}{\texttt{arXiv:2404.17377}}.

\bibitem{CMS:2011aa}
\hrefCMSnoop {}{{CMS Collaboration}, ``Measurements of inclusive
  $\uppercase{W}$and $\uppercase{Z}$ cross sections in pp collisions at $\sqrt
  {s}$ = 7 $\uppercase{T}$ev'',} \textit{ JHEP} \textbf{ 10} (2011) 132,
  \href{http://dx.doi.org/10.1007/JHEP10(2011)132}{\doi{10.1007/JHEP10(2011)132}},
  \href{http://www.arXiv.org/abs/1107.4789}{\texttt{arXiv:1107.4789}}.

\bibitem{Chatrchyan:2014fea}
\hrefCMSnoop {}{{CMS Collaboration}, ``{Description and performance of track
  and primary-vertex reconstruction with the CMS tracker}'',} \textit{ JINST}
  \textbf{ 9} (2014) P10009,
  \href{http://dx.doi.org/10.1088/1748-0221/9/10/P10009}{\doi{10.1088/1748-0221/9/10/P10009}},
\href{http://www.arXiv.org/abs/1405.6569}{\texttt{arXiv:1405.6569}}.

\bibitem{Chatrchyan:2012xi}
\hrefCMSnoop {}{{CMS Collaboration}, ``Performance of {CMS} muon reconstruction
  in pp collision events at {$\roots= 7\TeV$}'',} \textit{ JINST} \textbf{ 7}
  (2012) P10002,
  \href{http://dx.doi.org/10.1088/1748-0221/7/10/P10002}{\doi{10.1088/1748-0221/7/10/P10002}},
\href{http://www.arXiv.org/abs/1206.4071}{\texttt{arXiv:1206.4071}}.

\bibitem{Lokhtin:2011qq}
\hrefCMSnoop {}{I.~Lokhtin, A.~Belyaev, and A.~Snigirev, ``{Jet quenching
  pattern at LHC in PYQUEN model}'',} \textit{ Eur. Phys. J. C} \textbf{ 71}
  (2011) 1650,
  \href{http://dx.doi.org/10.1140/epjc/s10052-011-1650-1}{\doi{10.1140/epjc/s10052-011-1650-1}},
  \href{http://www.arXiv.org/abs/1103.1853}{\texttt{arXiv:1103.1853}}.

\bibitem{Zapp:2008gi}
K.~Zapp\hrefCMSnoop {}{ { et~al.}, ``{A Monte Carlo model for 'jet
  quenching'}'',} \textit{ Eur. Phys. J. C} \textbf{ 60} (2009) 617,
  \href{http://dx.doi.org/10.1140/epjc/s10052-009-0941-2}{\doi{10.1140/epjc/s10052-009-0941-2}},
  \href{http://www.arXiv.org/abs/0804.3568}{\texttt{arXiv:0804.3568}}.

\bibitem{Zapp:2012ak}
\hrefCMSnoop {}{K.~C. Zapp, F.~Krauss, and U.~A. Wiedemann, ``{A perturbative
  framework for jet quenching}'',} \textit{ JHEP} \textbf{ 03} (2013) 080,
  \href{http://dx.doi.org/10.1007/JHEP03(2013)080}{\doi{10.1007/JHEP03(2013)080}},
\href{http://www.arXiv.org/abs/1212.1599}{\texttt{arXiv:1212.1599}}.

\bibitem{Zapp:2013vla}
\hrefCMSnoop {}{K.~C. Zapp, ``{JEWEL 2.0.0: directions for use}'',} \textit{
  Eur. Phys. J. C} \textbf{ 74} (2014) 2762,
  \href{http://dx.doi.org/10.1140/epjc/s10052-014-2762-1}{\doi{10.1140/epjc/s10052-014-2762-1}},
  \href{http://www.arXiv.org/abs/1311.0048}{\texttt{arXiv:1311.0048}}.

\bibitem{KunnawalkamElayavalli:2017hxo}
\hrefCMSnoop {}{R.~Kunnawalkam~Elayavalli and K.~C. Zapp, ``{Medium response in
  JEWEL and its impact on jet shape observables in heavy ion collisions}'',}
  \textit{ JHEP} \textbf{ 07} (2017) 141,
  \href{http://dx.doi.org/10.1007/JHEP07(2017)141}{\doi{10.1007/JHEP07(2017)141}},
\href{http://www.arXiv.org/abs/1707.01539}{\texttt{arXiv:1707.01539}}.

\bibitem{Casalderrey-Solana:2020rsj}
J.~Casalderrey-Solana\hrefCMSnoop {}{ { et~al.}, ``{Jet wake from linearized
  hydrodynamics}'',} \textit{ JHEP} \textbf{ 05} (2021) 230,
  \href{http://dx.doi.org/10.1007/JHEP05(2021)230}{\doi{10.1007/JHEP05(2021)230}},
  \href{http://www.arXiv.org/abs/2010.01140}{\texttt{arXiv:2010.01140}}.

\bibitem{Chen:2017zte}
W.~Chen\hrefCMSnoop {}{ { et~al.}, ``{Effects of jet-induced medium excitation
  in $\gamma$-hadron correlation in A+A collisions}'',} \textit{ Phys. Lett. B}
  \textbf{ 777} (2018) 86,
  \href{http://dx.doi.org/10.1016/j.physletb.2017.12.015}{\doi{10.1016/j.physletb.2017.12.015}},
\href{http://www.arXiv.org/abs/1704.03648}{\texttt{arXiv:1704.03648}}.

\end{thebibliography}\endgroup

\cleardoublepage \appendix\section{The CMS Collaboration \label{app:collab}}\begin{sloppypar}\hyphenpenalty=5000\widowpenalty=500\clubpenalty=5000\cmsinstitute{Yerevan Physics Institute, Yerevan, Armenia}
{\tolerance=6000
V.~Chekhovsky, A.~Hayrapetyan, V.~Makarenko\cmsorcid{0000-0002-8406-8605}, A.~Tumasyan\cmsAuthorMark{1}\cmsorcid{0009-0000-0684-6742}
\par}
\cmsinstitute{Institut f\"{u}r Hochenergiephysik, Vienna, Austria}
{\tolerance=6000
W.~Adam\cmsorcid{0000-0001-9099-4341}, J.W.~Andrejkovic, L.~Benato\cmsorcid{0000-0001-5135-7489}, T.~Bergauer\cmsorcid{0000-0002-5786-0293}, S.~Chatterjee\cmsorcid{0000-0003-2660-0349}, K.~Damanakis\cmsorcid{0000-0001-5389-2872}, M.~Dragicevic\cmsorcid{0000-0003-1967-6783}, P.S.~Hussain\cmsorcid{0000-0002-4825-5278}, M.~Jeitler\cmsAuthorMark{2}\cmsorcid{0000-0002-5141-9560}, N.~Krammer\cmsorcid{0000-0002-0548-0985}, A.~Li\cmsorcid{0000-0002-4547-116X}, D.~Liko\cmsorcid{0000-0002-3380-473X}, I.~Mikulec\cmsorcid{0000-0003-0385-2746}, J.~Schieck\cmsAuthorMark{2}\cmsorcid{0000-0002-1058-8093}, R.~Sch\"{o}fbeck\cmsAuthorMark{2}\cmsorcid{0000-0002-2332-8784}, D.~Schwarz\cmsorcid{0000-0002-3821-7331}, M.~Sonawane\cmsorcid{0000-0003-0510-7010}, W.~Waltenberger\cmsorcid{0000-0002-6215-7228}, C.-E.~Wulz\cmsAuthorMark{2}\cmsorcid{0000-0001-9226-5812}
% ...
\par}
\vskip\cmsinstskip
\dag:~Deceased\\
$^{1}$Also at Yerevan State University, Yerevan, Armenia\\
$^{2}$Also at TU Wien, Vienna, Austria\\
% ...
\end{sloppypar}
\end{document}